\documentclass[twoside,english,superscriptaddress,pre,showkeys,aps,twocolumn]{revtex4-1}
\usepackage[T1]{fontenc}
\usepackage[latin9]{inputenc}
\usepackage{color}
\usepackage{amsthm}
\usepackage{amsmath}
\usepackage{amssymb}
\usepackage{graphicx}
\usepackage{esint}

\makeatletter

\@ifundefined{textcolor}{}
{%
\definecolor{BLACK}{gray}{0}
\definecolor{WHITE}{gray}{1}
\definecolor{RED}{rgb}{1,0,0}
\definecolor{GREEN}{rgb}{0,1,0}
\definecolor{BLUE}{rgb}{0,0,1}
\definecolor{CYAN}{cmyk}{1,0,0,0}
\definecolor{MAGENTA}{cmyk}{0,1,0,0}
\definecolor{YELLOW}{cmyk}{0,0,1,0}
}

\@ifundefined{showcaptionsetup}{}{%
 \PassOptionsToPackage{caption=false}{subfig}}
\usepackage{subfig}
\makeatother

\usepackage{babel}
\begin{document}
\captionsetup{justification=centerlast,font={footnotesize},position=below}
\renewcommand{\figurename}{FIG.} \addto\captionsenglish{\renewcommand{\figurename}{FIG.}}
\renewcommand{\thefigure}{\arabic{figure}}

\title{Positive association and global connectivity in dependent percolation}

\author{Navid Dianati}

\email{navid@umich.edu}

\thanks{corresponding author.}

\author{Yen Ting Lin}

\email{yentingl@umich.edu}

\affiliation{Department of Physics, University of Michigan, Ann Arbor, MI 48109-104}
\begin{abstract}
We study the effect of positive correlations on the critical threshold
of site and bond percolation in a square lattice with $d=2$. We propose
two algorithms for generating dependent lattices with minimal correlation
length and non-negative $n$-point correlations whose critical behavior
is then compared with that of independent lattices. For site percolation,
we show numerically that the introduction of this specific form of
positive correlation results in a lower percolation threshold, i.e.,
higher connectivity. For bond percolation, the opposite is observed.
In this case, however, we show that the dual lattice is also totally
positively associated, demonstrating that positive association can
result in either an increase or a decrease in global connectivity.
\end{abstract}


\maketitle

\section{Introduction}

Independent site and bond percolation have been the subject of extensive
studies in the past few decades. The critical thresholds of several
lattice geometries are known thanks to rigorous and non-rigorous arguments,
and the critical behavior for general classes of lattices at or near
criticality are well understood via scaling arguments \citep{stauffer1994introduction,GrimmettPercolation1989}. 

The problem of \textit{dependent percolation }on the other hand, is
relatively untouched due to the rapid growth of the mathematical complexity
as correlations between sites or bonds are introduced. In a series
of papers in the 1970s, Fortuin and Kasteleyn \citep{fortuin1972randomI,fortuin1972randomII,fortuin1972randomIII}
proved the equivalence between various Potts models and dependent
percolation with a probability measure different from the independent
product measure in that it favors more clusters. The link between
Potts models and percolation allowed results on universality and critical
exponents in the former to be applied to the latter. Regarding dependent
percolation for instance, it was shown \citep{weinrib1984long} that
long-range correlations fail to alter the universality class of a
percolation process provided the correlation function asymptotically
decays with distance rapidly enough. While the critical behavior of
dependent percolation is generally understood, however, the value
of the critical threshold in general depends on the specifics of the
probability measure; a relationship that remains obscure.

In a major development, Aizenman and Grimmett \citep{aizenman1991strict}
established sufficient conditions for the \textit{monotonicity\index{monotonicity}
}of the critical point in a particular family of dependent percolation
processes obtained from independent ones by means of an \textit{enhancement}:\index{Enhancements!}
the procedure whereby a new ensemble of lattice realizations is generated
by applying an optionally stochastic cellular automaton with local
rules to the realizations of the independent lattice resulting in
some previously closed sites (bonds) being declared open. 

It was shown that for \textit{essential enhancements}\index{Enhancements!Essential},
i.e., enhancements capable of creating a doubly infinite path for
at least one independent realization without such a path, the critical
point of the enhanced lattice is strictly lower than the independent
lattice. In other words, in an independent lattice, for any given
essential enhancement, there exists a range of sub-critical occupation
probabilities $\pi_{c}=(p_{0},p_{c})$ such that at $p\in\pi_{c},$
systematically adding open sites to realizations of the lattice according
to the enhancement will turn the process supercritical. Other dependent
models obtained by enhancements or other cellular automata have been
studied as well (e.g. \citep{camia2004particular,camia2008universality}). 

A related question is whether monotonicity in the critical threshold
(or lack thereof) can be deduced from the characterization of correlations
in the lattice. In particular, what can be said about the critical
threshold if all sites (bonds) are positively correlated?

Let us make this question precise. Let a finite lattice of bonds be
defined as the edge set $E=\{e_{i}\}$ of a graph $G=(V,E)$ and let
the sample space $\Omega=\left\{ 0,1\right\} ^{E}$ be the set of
all realizations. Let $\mathcal{F}=2^{\Omega},$ the power set of
$\Omega,$ be the set of all events and denote by $a(e_{i})\in\mathcal{F}$
the event that the edge $e_{i}$ be open. Similarly, for a site percolation
problem let $V=\{v_{i}\},$ $\Omega=\{0,1\}^{V},$ $\mathcal{F}=2^{\Omega}$
and $a(v_{i})\in\mathcal{F}$ denote the analogous quantities. Most
of our discussions are articulated in terms of the bond percolation
problem, but the obvious analogues exist for site percolation as well. 

Consider the following probability measures on the same lattice (graph):
\begin{enumerate}
\item $\Pr\left[a(e_{i})\right]=p_{i}$ for all $i,$ and all $a(e_{i})$
are independent.
\item $\Pr\left[a(e_{i})\right]=p_{i}$ for all $i,$ and $\Pr\left[\cap_{e_{i}\in E_{k}}a(e_{i})\right]\geq$\\
$\prod_{e_{i}\in E_{k}}\Pr(a(e_{i}))$ for all subsets $E_{k}\subset E.$
\end{enumerate}
In case (1), all edges are independent, whereas in case (2), all subsets
of edges are positively correlated if not independent. In both cases
though, the single-edge probabilities are given by the same set $\{p_{i}\}.$
This is a very particular construction, but as we will discuss below,
it arises naturally in important problems involving coarse-graining
of independent lattices.

We ask the following question: Is it true that if lattice (1) percolates
for a particular set of edge probabilities $\left\{ p_{i}\right\} ,$
then lattice (2) also percolates at the same point in the parameter
space? In other words, we set out to test the intuitive hypothesis
that introducing positive association across the board while keeping
everything else constant in fact increases global connectivity.

The rest of this paper is organized as follows: first, for each of
the two percolation models (site and bond), we propose an algorithm
by which a correlated lattice is derived from an un-correlated \textit{primal}
lattice. In both cases, each bond (site) is correlated only with its
nearest neighbors. We believe this to be the simplest possible model
of correlated percolation, with minimal additional control parameters
necessary. Next, we present simulation results for both models, computing
the critical point for each case. Finally, we demonstrate that common
arguments fail to predict the results and discuss the difficulties
in applying symmetry arguments to these problems.

\section{Dependent site percolation}

Beginning with site percolation, we consider the problem where the
only 2-point correlations exist between nearest neighbors. To be more
precise, we ask what happens to the critical point of a lattice if
uniform nearest-neighbor correlations are added while keeping the
single site occupancies unchanged. Naturally, one has to ask first
whether such setup is possible in the first place. Here, we present
a simple algorithm that given an uncorrelated lattice with occupation
probability $p,$ generates a second lattice with occupation probability
also equal to $p,$ but in which there is a uniform nearest neighbor
correlation between sites.

Consider a two-dimensional square lattice $V$ of sites indexed by
an ordered pair $(i,j)\in\mathbb{Z}^{2}$ and occupied with probability
$p$. The state of each site is represented by a random variable $s_{i,j}$
where $s_{i,j}=1$ if the site is occupied and $0$ otherwise. One
can think of the $s_{i,j}$ as the indicator functions
\begin{equation}
s_{i,j}=\boldsymbol{1}_{X_{i,j}<p}
\end{equation}
where $X_{i,j}\sim\mbox{Uniform(0,1) }$are identical independent
random variables.. We shall henceforth refer to the $X_{i,j}$ as
the underlying random variables of $V.$

Given a realization of such a lattice, we define a second lattice
$V'$ indexed by $\left(2\mathbb{Z}-1\right)^{2}$ and for every $(i,j)\in V',$
the state of the site is given by $u_{i,j}=\boldsymbol{1}_{Y_{i,j}<q}$
where $Y_{i,j}$ is a random variable defined as the arithmetic mean
of the nearest neighbors of the $(i,j)$ site in $V$: 
\begin{equation}
Y_{i,j}=\frac{1}{4}\left(X_{i+1,j}+X_{i-1,j}+X_{i,j+1}+X_{i,j-1}\right).
\end{equation}
Our goal is to choose $q$ such that $\Pr(u_{i,j}=1)=\Pr(s_{i,j}=1)=p.$
\begin{figure}[t]
\begin{centering}
\includegraphics[scale=0.8]{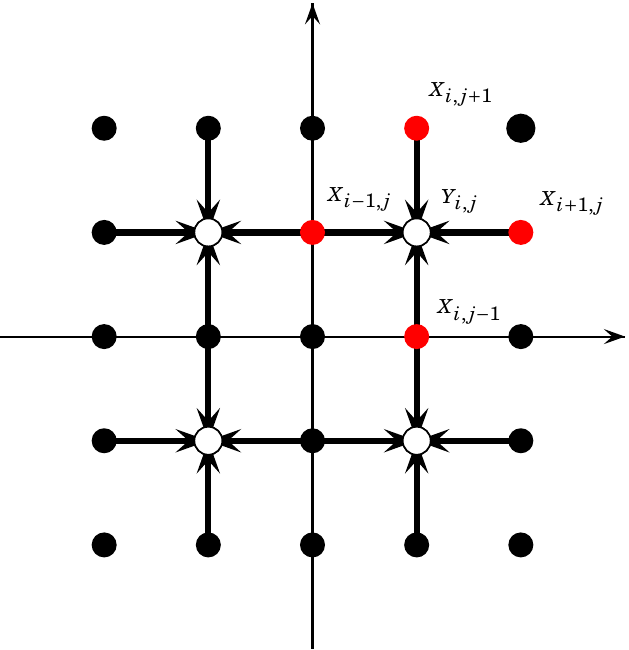}\caption{{\small The underlying random variables of the primal lattice (solid
dots) are averaged to generate those of the correlated lattice (empty
circles).}}

\par\end{centering}

\end{figure}
Clearly, each site in $V'$ is only correlated with its four nearest
neighbors. The correlation between neighboring $V'$ sites as well
as the threshold probability $q$ necessary to define the state of
each site remain to be determined.

In order to compute $q,$ we first derive an expression for the probability
density function of $Y_{i,j}.$ For the sake of readability, consider
one site in $V'$ with underlying random variable $Y$ and denote
its four $V$ neighbors by $X_{1},X_{2},X_{3},X_{4}.$ Define $\phi_{X}(t)$
to be the characteristic function of the $X$ variables (which are
identical): $\phi_{X}(t)=\int_{-\infty}^{\infty}e^{-itx}f_{X}(x)\, dx$
where $f_{X}(x)$ is the probability density function of $X_{i}:$
\begin{equation}
f_{X}(x)=\begin{cases}
1 & 0\leq x\leq1\\
0 & \mbox{otherwise}.
\end{cases}
\end{equation}
It is an elementary result that if the $X_{i}$ are independent, then
$\phi_{\sum X_{i}}(t)=\prod\phi_{X_{i}}(t)$ \citep{grimmett2001probability}.
Furthermore, by the inversion formula for the characteristic function,
the density function for $\sum_{i=1}^{4}X_{i}$ is given by $g(x)=\frac{1}{2\pi}\int_{-\infty}^{\infty}e^{itx}\phi_{\sum X_{i}}(t)\, dt$
\citep{grimmett2001probability}. Noting that the probability density
function of $Y=\frac{1}{4}\sum_{i=1}^{4}X_{i},$ $f_{Y}(y)$ is equal
to $4g(4y),$ a straightforward calculation yields:
\begin{align}
f_{Y}(y)=\frac{1}{3} & \left[|4y-4|^{3}-4|4y-3|^{3}\right.\notag\\
\left.+6|4y\right. & \left.-2|^{3}-4|4y-1|^{3}+64y^{3}\right]
\end{align}
for $0\leq y\leq1$ and zero otherwise. Then, the cumulative distribution
function of $Y$, $F_{Y}(y)=\int_{0}^{y}f_{Y}(z)\mathrm{d}z$ may
be directly computed.

\begin{figure}[t]
\centering{}\includegraphics[scale=0.6]{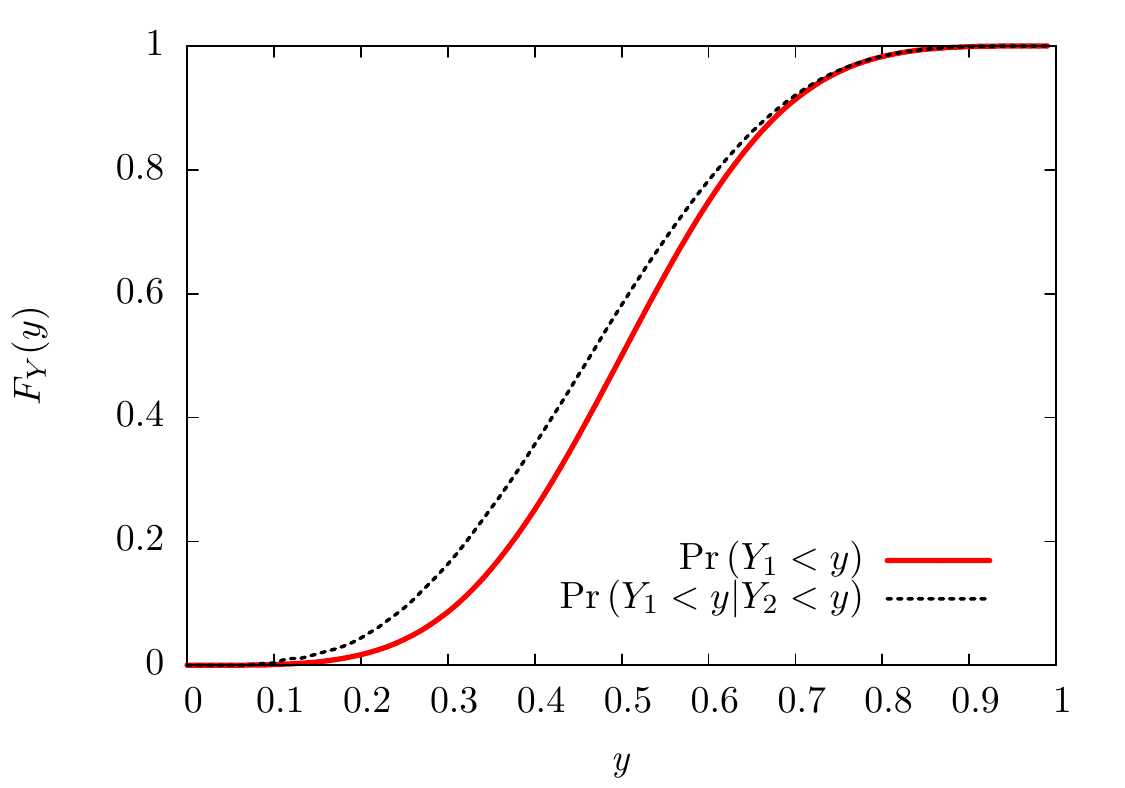}\caption{{\small The cumulative distribution function of a second layer random
variable $Y_{1}=\frac{1}{4}\left(X_{1}+X_{2}+X_{3}+X_{4}\right)$
and the conditional distribution function for two second-layer random
variables $Y_{1}$ and $Y_{2}=\frac{1}{4}\left(X_{1}+X_{5}+X_{6}+X_{7}\right).$
\label{fig:site-probabilities}}}
\end{figure}

Thus, with the sites in the $V'$ defined as $u_{i,j}=\boldsymbol{1}_{Y_{i,j}<q}$,
we have 
\begin{equation}
q=F_{Y}^{-1}(p)\Longrightarrow\Pr(u_{ij}=1)=p.
\end{equation}
That is, a dependent lattice populated with density $p$ must be derived
from a primal lattice populated with density $q=F_{Y}^{-1}(p)$ which
needs to be solved numerically. Next, we derive an expression for
the nearest-neighbor conditional occupation probability in this model.
For the sake of generality, we slightly modify the notation used above:
we denote the cumulative distribution function of the average of $n$
independent uniformly distributed random variables, $Y=\frac{1}{n}\sum_{i=1}^{n}X_{i}$
by $F^{[n]}(y).$ Consider two such random variables: $Y_{1}$ and
$Y_{2}$ defined by
\begin{align}
Y_{1}= & \frac{1}{n}\left(Z+U_{1}+U_{2}+\cdots+U_{n-1}\right),\\
Y_{2}= & \frac{1}{n}\left(Z+W_{1}+W_{2}+\cdots+W_{n-1}\right)
\end{align}
where $Z,U_{i},W_{i},i=1,\cdots,n-1$ are independently drawn from
$\mathrm{Uniform}(0,1).$ Then, $\Pr(Y_{1}<q|Y_{2}<q)=\Pr(Y_{1}<q\,\cap\, Y_{2}<q)\allowbreak/\Pr(Y_{2}<q).$
The numerator is\thickmuskip=0mu\thinmuskip=0mu
\begin{align}
 & \Pr(Y_{1}<q\cap Y_{2}<q)=\int_{0}^{1}\Pr(Y_{1}<q\cap Y_{2}<q|Z=s)f_{Z}(s)\mathrm{d}s\notag\\
 & =\int_{0}^{1}\left[\Pr\left(\sum_{i=1}^{n-1}U_{i}<nq-s\right)\Pr\left(\sum_{i=1}^{n-1}W_{i}<nq-s\right)f_{Z}(s)\right]\mathrm{d}s\notag\\
 & =\int_{0}^{1}\left[F^{[n-1]}\left(\frac{nq-s}{n-1}\right)\right]^{2}f_{Z}(s)\mathrm{d}s.
\end{align}
Thus,\thickmuskip=3mu\thinmuskip=3mu
\begin{align}
\Pr(Y_{1}<q|Y_{2}<q) & =\frac{\int_{0}^{1}\left[F^{[n-1]}\left(\frac{nq-s}{n-1}\right)\right]^{2}\mathrm{d}s}{F^{[n]}(q)}
\end{align}
This relation may be used to compute---numerically or in closed form---
the correlation between nearest neighbors in the model. Figure \ref{fig:site-probabilities}
shows $F^{[4]}(y)$ and the corresponding nearest-neighbor conditional
probability. That $Y_{1}$ and $Y_{2}$ are positively correlated
is evident.

\section{Dependent bond percolation}

\begin{figure}[t]
\noindent \subfloat[\label{fig:diagonal-lattice}]{\includegraphics[scale=0.42]{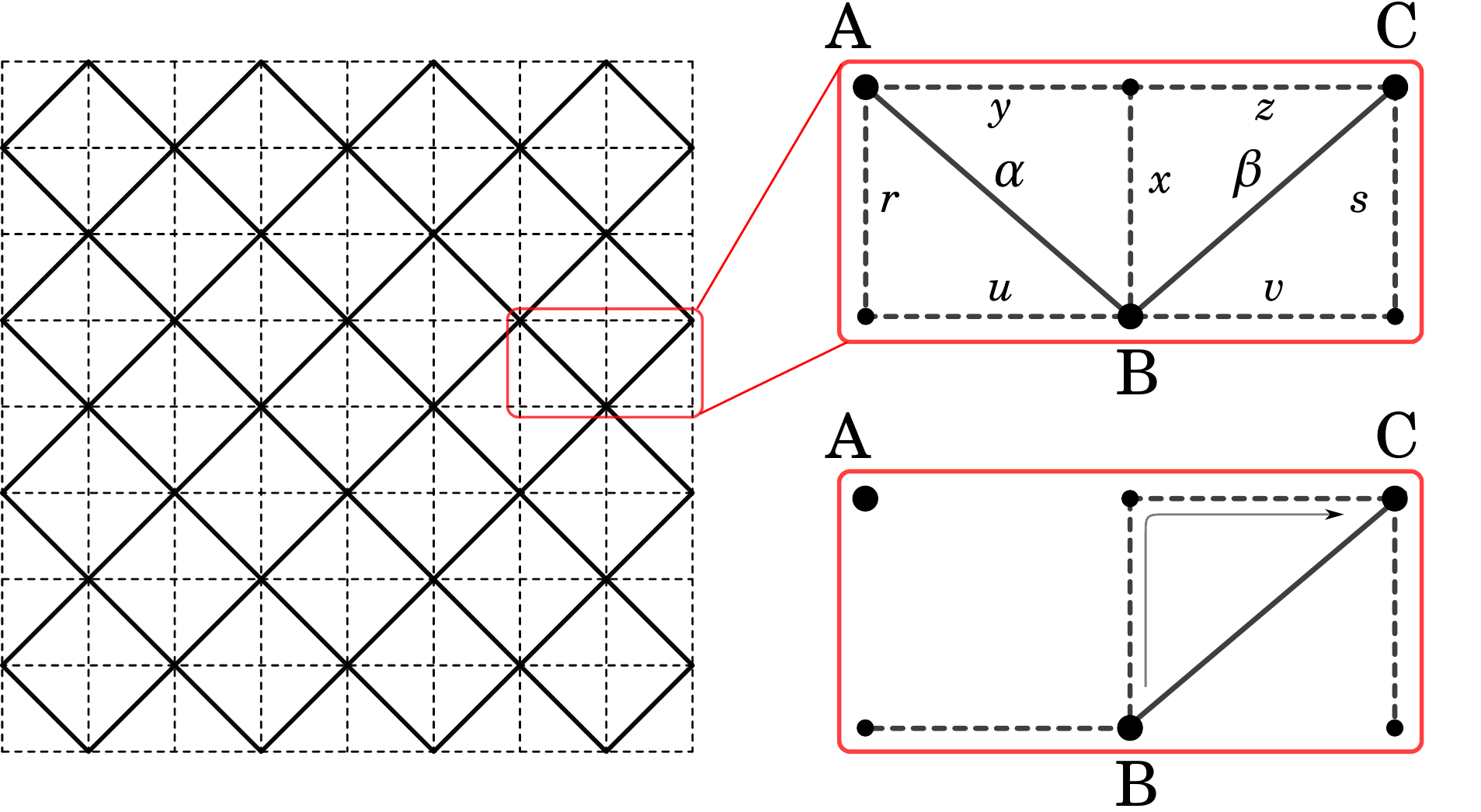}}\\
\subfloat[\label{fig:diagonal-lattice-generation}]{\centering{}\includegraphics[scale=0.43]{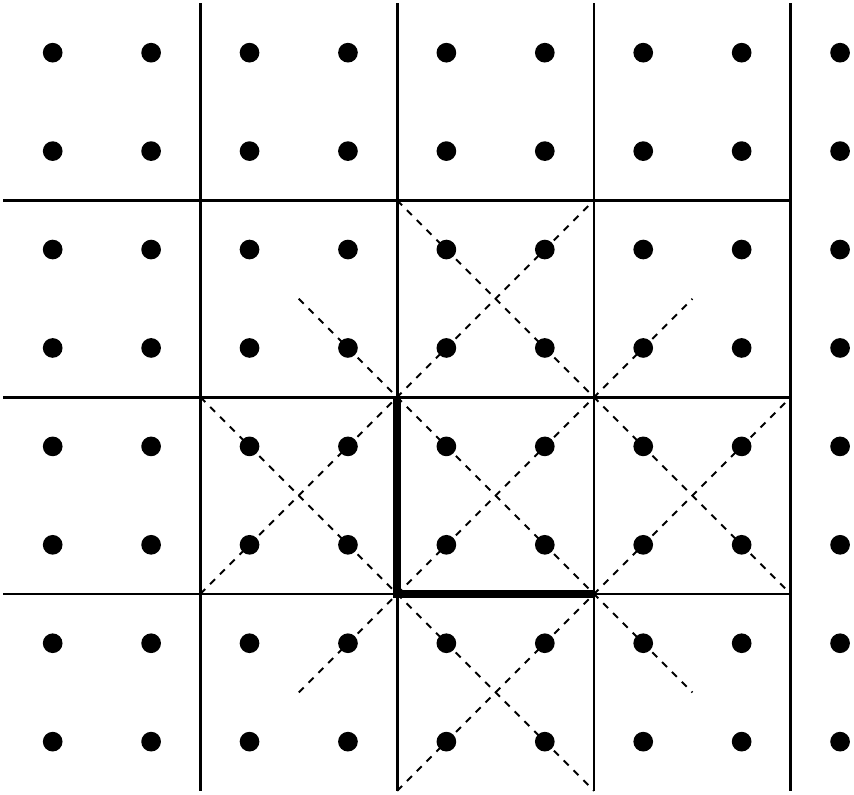}}~~~~~~\subfloat[]{\noindent \includegraphics[scale=0.17]{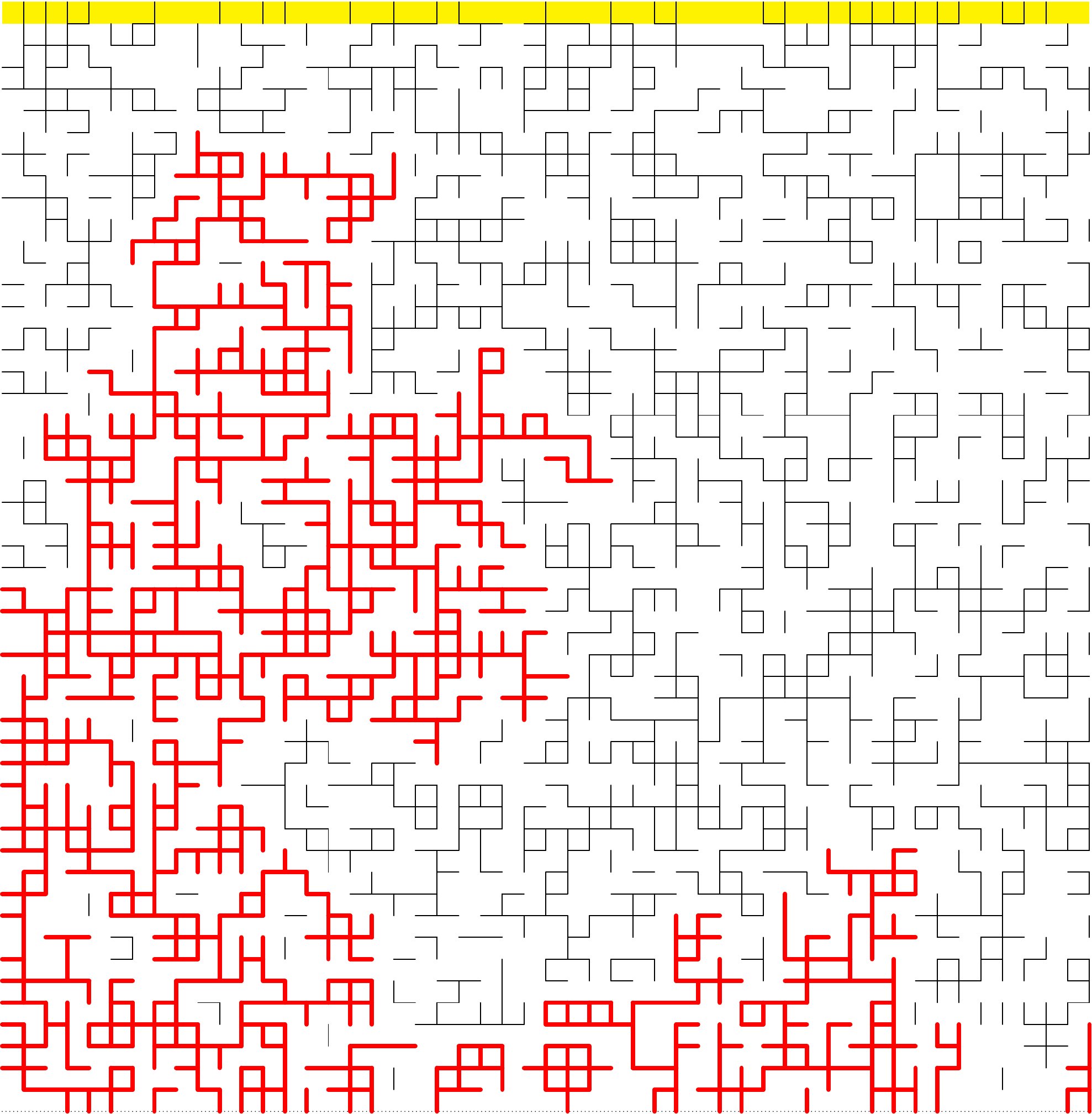}

\noindent }\caption{{\small (a) Renormalization scheme for bond percolation. The underlying
lattice is populated with identical and independent bonds: for $a\in\{x,y,z,u,v,w,r,s\},$
$\Pr(a)=p.$ The }\textit{\small diagonal }{\small lattice is then
derived by observing the diagonal connectivity within each square:
$\Pr(B\leftrightarrow C)=\Pr(x)\Pr(z)+\Pr(v)\Pr(s)-\Pr(x)\Pr(z)\Pr(v)\Pr(s)=2p^{2}-p^{4}.$
(b) Implementation of the renormalization in practice. (c) {[}color
online{]} A realization of dependent bond percolation.}}
\end{figure}

In order to illustrate a different application, we devise a different
strategy for generating derived lattices with positive nearest-neighbor
correlations in the case of bond percolation. Figure \ref{fig:diagonal-lattice}
illustrates a renormalization scheme where a coarse-grained version
of an independent square lattice is constructed as follows. First,
the primal lattice (dashed lines) is populated with independent and
identical bonds with occupation probability $p.$ Given any such configuration,
we then proceed to construct the corresponding configuration of the
\textit{diagonal lattice }(solid lines) by placing an open diagonal
bond between diagonally opposite corners of any given squares of the
primal lattice whenever there is an open path between the two corners
through the primal edges on the boundary of the square. For instance,
if we denote by $x,y,r,u$ respectively the events that the four boundary
edges of a primal square be open and by $\alpha$ the event that the
corners $A$ and $B$ are connected through those edges (Fig. \ref{fig:diagonal-lattice}),
then $\alpha=(x\cap y)\cup(r\cap u),$ and $\Pr(\alpha)$ is equal
to
\begin{align}
\Pr(\alpha) & =2p^{2}-p^{4}.
\end{align}
Percolation on the diagonal lattice, then, implies percolation on
the primal lattice. Thus, the critical thresholds of the diagonal
lattices of the primal lattice and its dual together yield upper and
lower bounds on the critical threshold of the primal lattice.

For the purpose of simulating percolation on such a lattice, we require
a square $N\times N$ diagonal lattice which may be generated easily
as illustrated in Fig. \ref{fig:diagonal-lattice-generation}. Here,
the solid lines again represent the diagonal bonds, but they are rotated
by 45 degrees. The independent bonds of the primal lattice are represented
by dots which are populated independently, whereupon each diagonal
bond is declared open depending on the state of the four dots surrounding
it as discussed above.

In the diagonal lattice, then, any two nearest neighbors (more precisely,
nearest neighbors at a right angle with respect to one another) are
positively correlated, as they are both \textcolor{black}{increasing
functions} of the primal bond they share. In Fig. \ref{fig:diagonal-lattice},
the events $A\leftrightarrow B$ and $B\leftrightarrow C$ are positively
correlated due to their mutual dependence on the increasing event
$x.$ 

We may compute the correlation as follows:
\begin{align}
\Pr(B\leftrightarrow C|B\leftrightarrow A) & =\Pr(\alpha|\beta)\notag\\
=\Pr(\alpha|\beta\cap x)\Pr & (x|\beta)+\Pr(\alpha|\beta\cap\bar{x})\Pr(\bar{x}|\beta)
\end{align}
Using $\Pr(x|\beta)=\Pr(\beta|x)\Pr(x)/\Pr(\beta)$ and $\Pr(\bar{x}|\beta)=\Pr(\beta|\bar{x})\allowbreak\Pr(\bar{x}))\allowbreak/\Pr(\beta)$,
a straightforward calculation leads to:
\begin{align}
\Pr(B\leftrightarrow C|B\leftrightarrow A)= & \frac{p\left(1+p-p^{2}\right)^{2}}{2-p^{2}}+\frac{p^{2}(1-p)}{2-p^{2}}
\end{align}
On the other hand, the dual of the correlated lattice has an occupation
probability $\Pr(B\nleftrightarrow A)=\Pr(\bar{\alpha})=1-2p^{2}+p^{4}$
and a nearest-neighbor conditional probability equal to
\begin{align}
\Pr( & B\nleftrightarrow C|B\nleftrightarrow A)=\Pr(\bar{\alpha}|\bar{\beta})\notag\\
 & =\frac{(1-p)^{3}(1+p)^{2}(1+p-p^{2})}{1-2p^{2}+p^{4}}.
\end{align}
\begin{figure}[t]
\begin{centering}
\subfloat[]{\centering{}\includegraphics[scale=0.6]{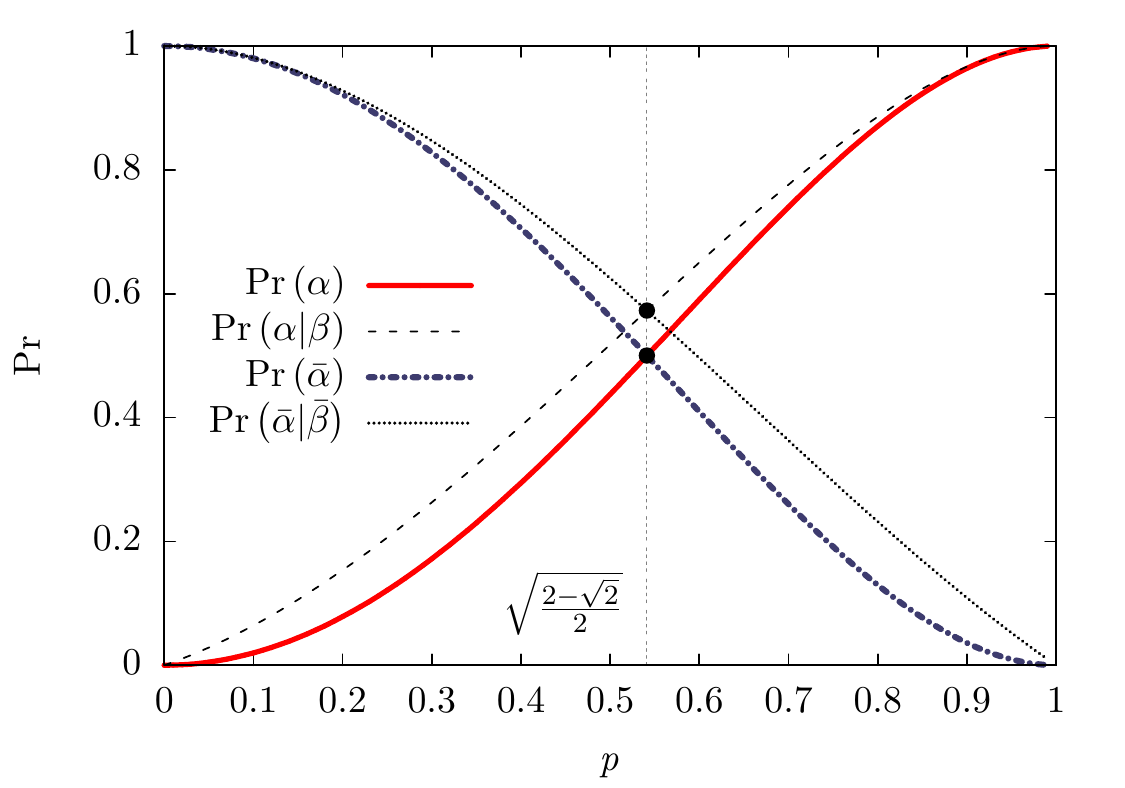}\-\-\-}
\par\end{centering}

\raggedleft{}\caption{{\small {[}color online{]} Bond percolation: the single-bond occupation
probability and the nearest-neighbor conditional occupation probability
for the diagonal lattice and its dual.}}
\end{figure}
Interestingly, one can verify that for $p\in[0,1],$
\begin{align}
\Big[\Pr(B\leftrightarrow C|B\leftrightarrow A)=\Pr(B\nleftrightarrow C|B\nleftrightarrow A)\Big]\notag\\
\Longleftrightarrow\Big[\Pr(B\nleftrightarrow A)=\Pr(B\leftrightarrow A)\Big].
\end{align}
When the diagonal lattice and its dual have equal occupation probabilities
of $0.5,$ they also have equal nearest-neighbor conditional occupation
probabilities. A simple calculation shows that this occurs at $p=\sqrt{\frac{1}{2}\left(2-\sqrt{2}\right)}$

\section{Simulations}

For each of the two models introduced above, we perform a set of simulations
to measure the critical occupation probability for various lattice
sizes $N.$ For each $N,$ we measure the crossing probability $\theta_{N}(p)$
for a large sample of occupation probabilities $p.$ Given the increasingly
sharp transition in $\theta_{N}(p)$ at the critical point, an adaptive
algorithm must be employed in order to generate a sample of $p$ values
concentrated about the unknown critical point. To this end, we implement
a Metropolis Monte-Carlo algorithm which stochastically samples the
set of $p$ values while attempting to minimize an ``energy function''
defined as an increasing function $E(\theta_{N}(p)-0.5)$ at a finite
but low ``temperature''.

\begin{figure}[t]
\centering{}\subfloat[{\small Site percolation.}]{\includegraphics[scale=0.5]{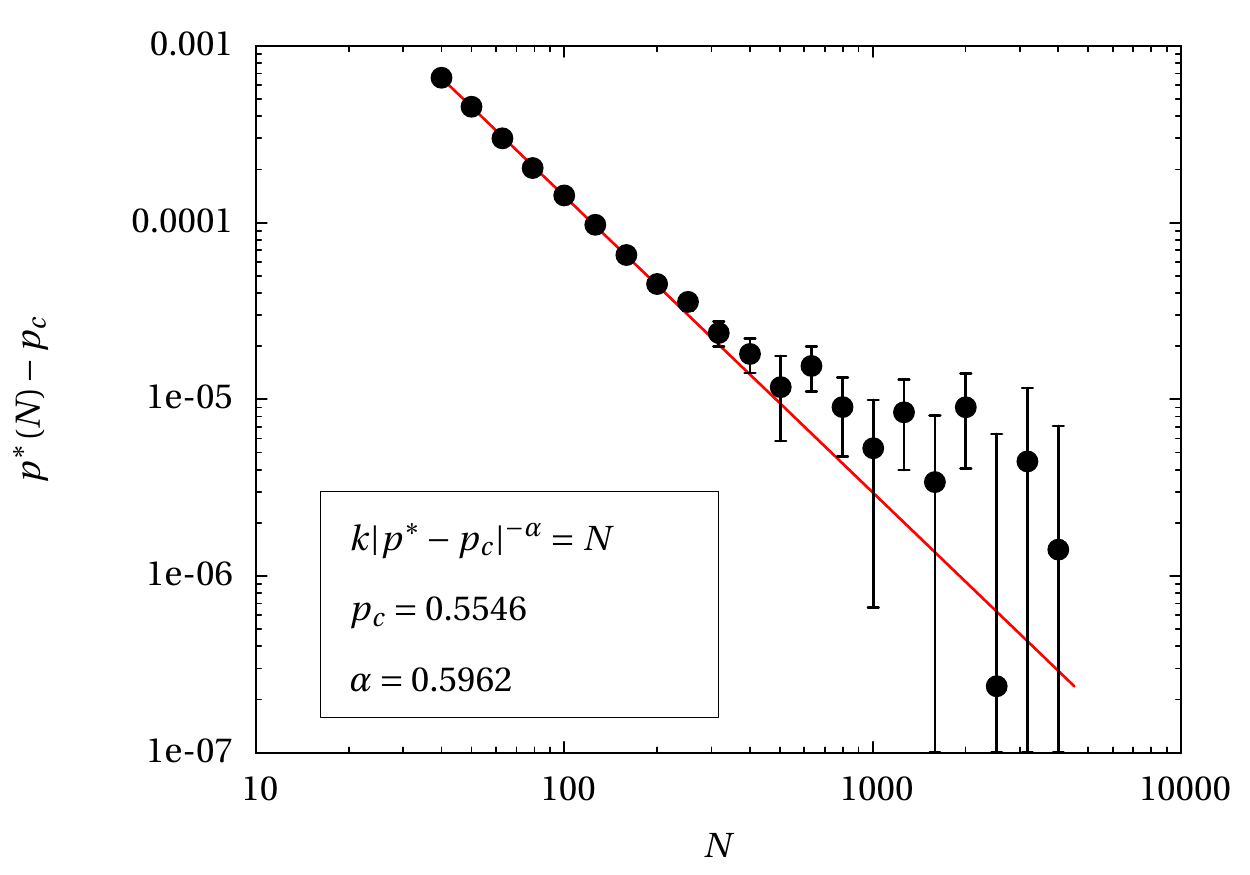}}\\
~\subfloat[Bond percolation.]{\includegraphics[scale=0.5]{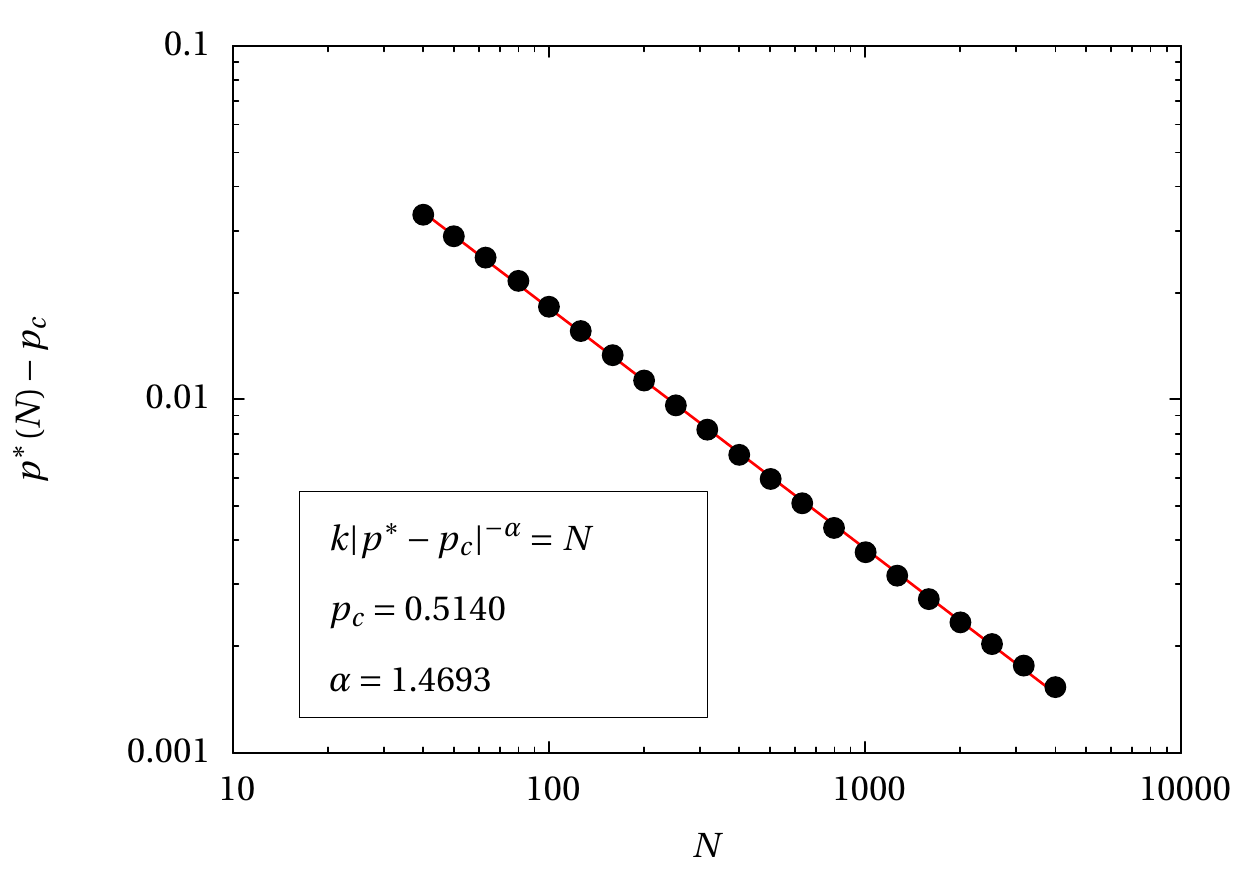}}\caption{{\small $p^{*}(N)$ as a function of lattice size $N.$ Dependent
percolation models.\label{fig:results-dependent-site-bond}}}
\end{figure}

For each $N,$ then, an optimal linear regression yields the finite-size
critical probability defined as $p^{*}(N)=\theta_{N}^{-1}(0.5)$,
and finally, the critical probability $p_{c}=\lim_{N\to\infty}p^{*}(N)$
is estimated by fitting a power-law to the set of finite-size critical
values:
\begin{equation}
k\left|p^{*}-p_{c}\right|^{-\alpha}=N.
\end{equation}
Figure \ref{fig:results-dependent-site-bond} shows our results for
dependent site and bond percolation. Whereas the critical probability
of independent percolation is roughly $0.5927,$ we find that the
dependent lattice percolates at $p_{c}\simeq0.5546,$ roughly $6\%$
lower.

For dependent bond percolation, we measure the critical threshold
to be $p_{c}\simeq0.5140$ or roughly a $3\%$ increase relative to
the independent threshold of $0.5$. As a benchmark, we also computed
the critical threshold for independent bond percolation and obtained
$p_{c}=0.49999.$

\section{Discussion}

Our results indicate a significant drop in the critical occupation
probability of site percolation with the introduction of positive
nearest-neighbor correlation, whereas the opposite occurs in bond
percolation. While the observed behavior in former seems compatible
with intuitive arguments (see below), a number of additional subtleties
are involved in the latter case, and new questions arise.

The duality argument ---used to prove that $p_{c}=0.5$ for independent
bond percolation~---may be roughly summarized as follows: the 2-D
square lattice is self-dual, and thus, at $p=0.5,$ the ensembles
of lattice realizations of the lattice and its dual are statistically
identical. Furthermore, the lattice percolates if and only if the
dual lattice does not (except perhaps at the critical point). Thus,
if $p_{c}$ is anything but the point where the two are statistically
identical, in this case $p=0.5,$ we face a contradiction. 

An extension of this argument may appear to apply to our dependent
model: now, the problem is parametrized not just by one parameter
(occupation probability), but a tuple (occupation probability and
nearest neighbor conditional occupation probability). There is no
correlation between a given bond and any other non-neighbor bond.
As we have seen above, when the occupation probabilities of the lattice
and its dual are equal (0.5), the conditional probabilities and thus
the nearest-neighbor correlations are also equal. At this point, the
lattice and its dual appear to be statistically identical. This is
not the case, however: while the only two-point correlation exists
between nearest neighbors, that alone is enough to generate non-trivial
$n$-point correlations for $n=3,4,5,\cdots.$ 

A different line of argument seeks to establish inequalities between
the dependent and independent lattices. The definition of the dependent
lattice based on an independent primal lattice allows us to immediately
apply useful correlation results from probability theory. In particular,
at any finite size, the primal lattice is trivially an associative
lattice: it satisfies the strong FKG condition and the FKG inequality
(see \citep{fortuin1971correlation} or for a modern presentation,
\citep{GrimmettPercolation1989}) applies. Further, with the obvious
partial ordering defined on the set of lattice realizations, clearly,
for each subset of the diagonal bonds, the event that those bonds
are occupied is an increasing event. Similarly, the event that they
are unoccupied is a decreasing event. 

Let $A_{k}$ be the event that all the edges in a subset $E_{k}\subset E$
be open. i.e., $A_{k}=\cap_{i}a(e_{i}).$ Any path between two points
$x,y$ in the lattice (or any cluster in general) corresponds to one
such event $A_{k}.$ The event $A_{k}$ will occur in each of the
two cases (1) and (2) with the following probabilities:
\begin{equation}
\Pr(A_{k})=\begin{cases}
\prod_{e_{i}\in E_{k}}\Pr\left(a(e_{i})\right) & \mbox{case 1}\\
\Pr\left(\bigcap_{e_{i}\in E_{k}}a(e_{i})\right) & \mbox{case 2}
\end{cases}
\end{equation}
The FKG inequality states that for a set $\{x_{i}\}$ of increasing
events (decreasing events), $\Pr(\bigcap x_{i})\allowbreak\geq\prod\Pr(x_{i}).$
We now apply this inequality to $\{a(e_{i})|e_{i}\in A_{k}\}$ for
every $k$, noting that $a(e_{i})$ are increasing events:
\begin{equation}
\Pr\left(\bigcap_{e_{i}\in E_{k}}a(e_{i})\right)\geq\prod_{e_{i}\in E_{k}}\Pr\left(a(e_{i})\right).\label{eq:FKG-result}
\end{equation}
 The result is that each path (or each cluster in general) is more
likely to be open in the presence of correlations than without correlations.
Note that the same inequality holds if we replace the increasing events
$a(e_{i})$ with the decreasing events $\overline{a}(e_{i}),$ meaning
that each path or cluster of the dual lattice also occurs with higher
probability once our correlations are introduced.

If we define percolation as the almost certain connectivity of opposite
boundaries of $B(N)$ (box of size $2N$, centered at the origin),
then as $N\to\infty,$ this may seem to imply that If the uncorrelated
lattice percolates at a point in the parameter space of single edge
probabilities, then the lattice will still percolate even if we introduce
positive correlations between edges, as long as we remain at the same
point in the single edge parameter space. This means that the supercritical
zone of the parameter space with correlations is a superset of the
supercritical zone of the parameter space without correlations. This
is true for the lattice as well as its dual, but the supercritical
zone of the lattice is the complement of the supercritical zone of
the dual lattice, which leads to the conclusion that the border between
the two--i.e, the critical surface-- remains intact.

However, to arrive at that conclusion, one must justify one more proposition,
namely that the \emph{union }of all the paths considered above also
increases in measure as a result of the introduction of the correlations.
As it turns out, this does not follow from the above argument.

This is a very peculiar situation. Every single path is more likely
to be open in the presence of correlations. However, the probability
of at least one of them being open is not guaranteed to increase. 

Similarly, one may attempt to show that the average cluster size does
not decrease when positive correlations are introduced, by showing
that each cluster $E_{k}\subset E$ appears with a higher probability
in that case. We have the same inequality as before, but with the
following caveat. While the probability of any cluster being open
$(\Pr\left[\cap_{e_{i}\in E_{k}}a(e_{i})\right]$) increases by introducing
positive correlations, the probability of occurrence of the \emph{lattice
animal}  made up of the same edges does not necessarily increase,
since the lattice animal consists of open interior edges \emph{and
}closed perimeter edges and the intersection of the two groups of
events (open interior edges and closed perimeter edges) does not necessarily
increase in measure. If we were able to prove an increase in the probability
of occurrence of all lattice animals, then the result would follow
trivially, but our situation is more complicated.

\section{Conclusions}

Our results allow us to answer the question posed at the outset. Using
our algorithm, we have simultaneously constructed two different dependent
bond percolation problems, on the lattice and on its dual. In both
cases, any subset of bonds is positively correlated. However, one
of the two has a lower critical point than the independent lattice
while (consequently), the other has a higher critical point. We see,
then, that positive correlation is not sufficient for increased connectivity,
nor is it sufficient for decreased connectivity.

\bibliographystyle{unsrtnat}
\bibliography{all}

\end{document}